\documentstyle[12pt]{article}
\begin{document}
\title{Quark and Pole Models of Nonleptonic Decays of Charmed
Baryons}
\author{P.\.{Z}enczykowski
\\H. Niewodnicza\'nski Institute of Nuclear Physics
\\ul. Radzikowskiego 152, 31-342 Krak\'ow, Poland
}
\maketitle
\begin{abstract}
Quark and pole models of nonleptonic decays of charmed baryons are analysed
from the point of view of their symmetry properties. The symmetry structure
of the parity conserving amplitudes that corresponds to the contribution of
the ground-state intermediate baryons is shown to differ from the one hitherto
employed in the symmetry approach. It is pointed out that the "subtraction"
of sea quark effects in hyperon decays leads to an estimate of $W$-exchange
contributions in charmed baryon decays that is significantly smaller than
naively expected
on the basis of $SU(4)$. An $SU(2)_{W}$ constraint questioning the reliability
of the factorization technique is exhibited. Finally, a successful fit to the
available data is presented.
\end{abstract}

\newpage
\section{Introduction}
Over the last ten years a description of heavy meson weak decays known as the
factorization approach has become available. In the arena of heavy baryon
decay theoretical progress has been fairly slow, however. To some extent this
state of affairs was conditioned by the type and quality of experimental data.
Only recently higher statistics 
data on nonleptonic weak decays of $\Lambda ^{+}_{c}$ have become available.
With several experiments on charmed baryons being now carried out at DESY,
Cornell, CERN and Fermilab the expected growth of the data basis has already
started to stimulate a more intensive theoretical effort in this area.

Although views have been expressed that the dynamics of nonleptonic weak decays
should become simpler as the decaying quark becomes heavier, a reliable
approach to the decays of charmed baryons does not exist thus far.
This is hardly surprising in view of the fact that in the long studied related
field of nonleptonic hyperon decays,
there is still no consensus as to
the relative importance and symmetry structure of various possible
contributions to both the parity violating and parity conserving amplitudes
\cite{DGH86}. In fact not only it is not clear there what is the relative size
and sign of pole model contributions from various intermediate
states
(i.e. mesons, ground-state and excited baryons - compare refs.
\cite{Ors79,Bon84,Nar88}) but even the value of the $f/d$
ratio
characterising the $SU(3)$ structure of the soft pion contribution to the
parity violating amplitudes is not agreed upon. The
valence
quark model predicts $f/d = -1$ while the phenomenological analysis of Pham
\cite{Pha84} suggests $f/d = -1.6$, much closer to that needed for a proper
description of the parity conserving amplitudes. Specific models to explain
such
deviation from the valence quark model have been proposed \cite{Don77,Zen92}.

Given this situation it seems unlikely that in the near future we shall be able
to predict through a reliable calculation the corresponding contributions in
the decays of charmed baryons. Instead, it is probably the incoming data that
will be instrumental in the broadening of our understanding of nonleptonic
weak decays for baryons in general and its deepening for hyperons in
particular.

To help resolve various emerging
                questions in a phenomenological way we adopt a
framework based on symmetry considerations. The main topic of this paper
is the discussion of the implications of various assumptions involving
and/or affecting symmetry properties of
models of nonleptonic decays of charmed baryons. The symmetry/quark model
approach adopted here is based on papers \cite{DDH80,Zen89}
and constitutes their generalization to the charmed baryon sector. Our
approach (briefly described in Section 2), although similar in spirit to the
one used previously in this context
\cite
{KKW79,KK92,KS91}, differs from the latter in an essential way. Namely, it
turns out that the symmetry
structure of the parity conserving amplitudes of refs \cite{KKW79,KK92}
does not correspond to that expected in the pole model with
ground-state baryons in the intermediate state.
In the present paper
the correct symmetry structure of the pole model with ground-state
intermediate baryons is used.
Thus, our paper essentially replaces the previous
symmetry-based papers on
charmed baryon decays. Apart from the above difference in
the treatment of parity conserving amplitudes
                    our paper differs from refs~\cite{KKW79,KK92} by a
more
phenomenological treatment of single-quark processes. Furthermore,
we  point out a couple of uncertainties and corrections hitherto  not
noticed in the literature (Section~\ref{sym-require}).
Finally, using symmetry approach as our framework we fit the existing
experimental data (Section~\ref{fits}).

\section{The Basic Quark/Pole Model}
\label{basic model}
The aim of this paper is to discuss
the symmetry structure of the quark and pole models
           of nonleptonic decays of charmed baryons and the implications of
various assumptions involving and/or affecting symmetry properties.
                     These assumptions may be tested
by comparing symmetry properties of their predictions for
the partial decay widths and asymmetries with the experimental ones.

For the decays with the emission of pseudoscalar mesons these partial
decay widths and asymmetries are given in terms of the parity violating
($A_{P}$) and the parity conserving ($B_{P}$) amplitudes by
\begin{eqnarray}
  \Gamma & = & \frac{1}{4\pi} \frac{k(E_{f}+m_{f})}{m_{i}}
           (|A_{P}|^{2}+|\bar{B}_{P}|^{2}) \nonumber \\
  \alpha & = & \frac{2 A_{P} \bar{B}_{P}}{|A_{P}|^{2}+|\bar{B}_{P}|^{2}}
  \label{eq:b01}
\end{eqnarray}
where
\begin{equation}
  \bar{B}_{P} = \frac{k}{E_{f}+m_{f}} B_{P}
\end{equation}
In Eq.(\ref{eq:b01}) $m_{i}$, $m_{f}$ are the masses of the initial and
final baryon, $E_{f}$ is the energy of the final baryon and $k$ its
decay momentum.

For the decays with the emission of vector mesons the corresponding
formulas read:

\begin{eqnarray}
 \Gamma & = & \frac{1}{4\pi } \frac{k(E_{f}+m_{f})}{m_{i}}
 (|A_{V_{\perp }}|^{2}+|\bar{B}_{V_{\perp }}|^{2}+
 |A_{V_{\parallel }}|^{2}+|\bar{B}_{V_{\parallel }}|^{2})  \nonumber \\
 \alpha & = & \frac{2 (A_{V_{\perp }} \bar{B}_{V_{\perp }} +
 A_{V_{\parallel }} \bar{B}_{V_{\parallel }}}
{(|A_{V_{\perp }}|^{2}+|\bar{B}_{V_{\perp }}|^{2}+
 |A_{V_{\parallel }}|^{2}+|\bar{B}_{V_{\parallel }}|^{2})}
\label{eq:b02}
\end{eqnarray}

where
\begin{equation}
 \bar{B}_{V_{\perp ,\parallel }} = \frac {k}{E_{f}+m_{f}}
      B_{V_{\perp ,\parallel }}
\end{equation}

and $A_{V_{\perp ,\parallel }}$ etc. are the amplitudes for the emission
of transverse ($\perp $) and longitudinal ($\parallel $) vector mesons.

The approach of this paper constitutes an application to the charmed
baryon sector of the quark-model technique used in the description of
the
$\Delta S~=~1$ hyperon decays in refs.~\cite{DDH80,Zen89,Bal67}. The main
idea of these papers was to separate from the dynamics the quark-model-based
spin-flavour symmetry relations between the amplitudes. The basic
reason for adopting such an approach was the lack of general consensus
among theorists on the relative size and sign of various dynamical
contributions. The approach of refs.~\cite{DDH80,Zen89} evades such
theoretical uncertainties by lumping various contributions into a few
reduced matrix elements to be fitted from experiment. These reduced
matrix elements correspond to quark diagrams of Fig.1. The diagrams
$(a)$, $(a')$ correspond to the so-called factorization amplitudes,
$(b1)$, $(b2)$ and $(d)$ are $W$-exchange contributions while
diagrams $(c1)$ and $(c2)$ summarize the effect of quark sea.
In fact,
K\"{o}rner  and his collaborators (see also ref.
\cite{Kor72}) never consider diagrams $(c)$, the inclusion of which is
crucial \cite{DDH80} for a proper description of hyperon decays.
(An important difference between the hyperon and
charmed-baryon sector is the absence of diagrams $(c)$ in the latter.
The implications of this difference shall be discussed in
Section 3.)

\subsection{Parity Violating Amplitudes}
\label{pv ampl}
Our approach to the parity violating amplitudes does not differ in an
essential way from that of refs.~\cite{KKW79,KK92,KS91}. The contribution
from the diagram $(d)$ is zero. Furthermore if $SU(4)$ symmetry were
exact    diagrams $(a)$ and $(a')$
                      would not contribute to the charmed baryon decays
with pseudoscalar meson emission (see e.g. ref.~\cite{DDH80}). For the
transverse vector mesons they do contribute, however, even in the limit
of exact $SU(4)$.

Calculations of the spin-flavour factors corresponding to the parity
violating amplitudes
$A_{P}$, $A_{V}$ were done using the quark model
technique of ref.~\cite{DDH80}. The results are gathered in Tables
1 and 2.

The reduced matrix elements $\tilde{a}$, $\tilde{a'}$ and $\tilde{b}$
are related to those of hyperon decays by

\begin{eqnarray}
\label{eq:b1}
\tilde {b} & = & b \cot \theta _{c} \nonumber \\
\tilde {a} & = & a \cot \theta _{c} \nonumber \\
\tilde{a'} & = & a' \cot \theta _{c}
\end{eqnarray}

where $\theta _{c}$ is the Cabibbo angle and the parameters $a$, $a'$,
$b$ are the corresponding reduced matrix elements for hyperon decays.
Their numerical values have been estimated in
refs.~\cite{DDH80,Zen89,Zen91} to be (in units of $10^{-7}$):

\begin{eqnarray}
\label{eq:b2a}
b & = & -5.0 \nonumber \\
a & = & +3.8 \nonumber \\
a'& = & -3.0
\end{eqnarray}

and, consequently, we have

\begin{eqnarray}
\label{eq:b2b}
\tilde{b} & = & -22.2 \nonumber \\
\tilde{a} & = & +16.9 \nonumber \\
\tilde{a'} & = & -13.3
\end{eqnarray}

In addition, the reduced matrix elements corresponding to the emission of
longitudinal ($\tilde{b'}$) and transverse ($\tilde{b}$) vector mesons are
related in the quark model by

\begin{equation}
\label{eq:b2c}
\tilde{b'} = \tilde{b}
\end{equation}

In the $SU(4)$-broken world, as the factorization approximation
indicates, diagrams $(a)$ and $(a')$ do contribute to the decays
with pseudoscalar meson emission. This has been taken into account
in Table 1 where such contributions have been given strength $g$ and
$g'$ respectively.

Estimates of $g$ and $g'$ through factorization \cite{XK92,CT92}
give for both of them similar values (though with opposite signs)
of around (in units of $10^{-7}$)

\begin{equation}
\label{eq:b3}
4.0 - 6.0
\end{equation}

Comparing Eq.(\ref{eq:b2b}) with Eq.(\ref{eq:b3}) we see that for
charmed baryon decays - as for hyperon decays - the factorization
amplitudes still appear to give bigger contributions in the
$B~\rightarrow~B'V$ than in the $B~\rightarrow~B'P$ decays
although in the latter they are no longer negligible. We shall
come back to the discussion of the factorization amplitudes in
Subsections~3.3 and~\ref{factor sextet}.

The nonnegligible size of the factorization amplitudes $g$, $g'$
- as required by nonvanishing experimental asymmetry in the
$\Lambda _{c}^{+}~\rightarrow~\Lambda ~\pi ^{+}$ decay - indicates
significant $SU(4)$ breaking effects resulting from large mass
difference between charmed and noncharmed (constituent) quarks:
\begin{equation}
\label{eq:b4}
m_{c}-m_{u,d,s} \approx 1.1 GeV
\end{equation}

Such a large mass difference must lead to significant differences
between the standard current algebra/quark model approach and the
pole model. In the pole model the dominant contribution to the
parity violating amplitudes comes from the lowest lying negative
parity $\frac{1}{2}^{-}$ excited baryons propagating between the
$W$-exchange and strong decay interactions shown in diagrams $(b1)$
and $(b2)$ of Fig.1. As discussed in ref.\cite{CT92,LaZen} the current
algebra and the pole model become equivalent in the $SU(4)$ limit
when

\begin{equation}
\label{eq:b5}
0 \leftarrow m_{c} - m_{u,d,s} \ll m(\frac{1}{2}^{-}) -
m(\frac{1}{2}^{+}) \approx 0.5 GeV
\end{equation}

Then, one can sum the contributions from the intermediate $\frac
{1}{2}^{-}$ baryon resonances and obtain the standard quark
model/current algebra prescription in which no information on the
intermediate $\frac{1}{2}^{-}$ states is needed.

In reality Eq.(\ref{eq:b5}) is of course not satisfied and significant
departures from simple current algebra predictions may be anticipated.
Such effects were discussed in ref.~\cite{CT92}. In this paper they
are not considered. The reasons behind their neglect are as follows.

First, we want to give a symmetry prediction that - unlike the one
given by K\"{o}rner and collaborators \cite{KKW79,KK92,KS91} - does
correspond to the standard pole model prescriptions for the parity
conserving amplitudes. Second, we want to point out other ambiguities
that as yet have not been discussed in the literature at all. Third,
we think that a reliable inclusion of $SU(4)$ breaking effects
might be very difficult. We believe that it will be the experiment
that will guide us on our way to a theoretical understanding of
how to properly take such effects into account. Accordingly, the
simple current algebra/quark model approach and its predictions
are of great interest themselves since they provide the basis for
future discussion of various departures from such simple models.

\subsection{Parity Conserving Amplitudes}
\label{pc ampl}
Calculation of the parity conserving amplitudes $B_{P}$, $B_{V}$ is
similar to the calculation of the previous subsection. There are two
main contributions to the amplitudes. The first is due to the
intermediate baryons (diagrams $(b1)$, $(b2)$, $(d)$), the second
(due to meson poles) is often treated in the factorization
approximation (diagrams $(a)$ and $(a')$). Evaluation of the
symmetry structure of the second contribution is straightforward
and leads to the pattern exhibited in Tables 1,2. In these tables
the reduced matrix elements corresponding to diagrams $(a)$ and
$(a')$ are denoted by $M$,$M'$ when the emitted meson is an
$SU(2)_{W}$ triplet ($P$,$V_{\perp }$) and by $m$,$m'$ when it is
an $SU(2)_{W}$ singlet ($V_{\parallel }$). In the next Section
we shall discuss these contributions and their actual size in more
detail.

The contribution from the intermediate baryons requires the calculation
of the spin-flavour structure of diagrams $(b1)$, $(b2)$, $(d)$. For
the $B_{c}~\rightarrow~BP$ decays the individual spin-flavour factors
corresponding to diagrams $(b1)$ and $(b2)$ are shown in Table 1.
In order to obtain the symmetry structure of the baryon pole
contributions to the parity conserving amplitudes, the spin-flavour
factors  corresponding to diagrams $(b1)$ and $(b2)$ have to be
multiplied by appropriate energy denominators and then added.
A closer inspection of these (assume $SU(3)$ i.e.: $\Sigma = \Lambda
= N = \Xi$ ; $ \Xi ^{+}_{c} = \Xi ^{0}_{c} = \Lambda ^{+}_{c}$)
               shows that if the intermediate baryons are in the
ground-state this addition procedure effectively results in the
subtraction of the spin-flavour factors of diagrams $(b1)$ and $(b2)$.
The same procedure, when applied to two versions of diagram $(d)$
(with $W$-exchange followed by strong decay and vice versa), leads
to the cancellation of these two contributions on account of their
identical spin-flavour structure. Thus, no overall contribution from
diagram $(d)$ is obtained. The above subtraction procedure may be
verified by explicitly calculating all the necessary $B'BM$ strong
couplings and weak baryon-to-baryon matrix elements
$<~B~|~H_{weak}^{p.c.}~|~B'~>$ and then combining them according to the
prescriptions of the pole model. In the process, the contributions
from $W$-exchanges between quarks not involved in meson emission
get cancelled and the symmetry structure of the resulting amplitudes
is that obtained from the subtraction of diagrams $(b1)$ and $(b2)$
(see also Appendix A of ref.~\cite{Zen92}).

If simple symmetry arguments are applied to link (the $W$-exchange
contribution to) the parity conserving
hyperon and charmed-baryon decay amplitudes one obtains for the
reduced matrix element $B$ of Tables 1 and 2 the value

\begin{equation}
\label{eq:b6}
B = 12 ( 1 - \frac{F}{D} ) C \cot \theta _{c}
\end{equation}

where $C = -33$ is the value fitted in hyperon nonleptonic decays
\cite{Zen89,Zen91} and $ F/D = 2/3$. In deriving Eq.(\ref{eq:b6})
we took into account the effect discussed in Subsection~\ref{SU(4)link}
which diminishes the size of $H^{p.c.}$ matrix elements in the charmed
baryon sector.

The estimate of Eq.(\ref{eq:b6}) is not correct, however, since it does
not consider the large difference in the size of pole factors
$1/(B_{i,f}-B')$ appearing in hyperon and charmed baryon parity
conserving amplitudes. More properly, Eq.(\ref{eq:b6}) should be replaced
by

\begin{equation}
\label{eq:b7}
B = 12 ( 1 - \frac{F}{D} ) C \cot \theta_{c} \frac{m_{\Sigma ,\Lambda}
- m_{N}}{m_{B_{c}} - m_{\Sigma ,\Lambda}}.
\end{equation}

Eq.(\ref{eq:b7}) gives as a rough estimate (in units of $10^{-7}$)

\begin{equation}
\label{eq:b8}
B \approx -95
\end{equation}

which compares well with the value -73 of ref.~\cite{XK92}.
Quark model relates the reduced matrix elements $B'$ and $B$ by

\begin{equation}
\label{eq:b9}
B' = B
\end{equation}

We proceed now to the discussion of the implications of various assumptions
affecting and/or involving symmetry properties.

\section{Discussion}
\label{sym-require}
\subsection{Symmetry Structure of Parity Conserving Amplitudes}
\label{subtraction}
The symmetry structure of the parity conserving amplitudes in the
standard pole model differs from that given by K\"{o}rner and
collaborators \cite{KKW79,KK92,KS91}. In the pole model of
Section (\ref{pc ampl}) flavour symmetry is kept in strong vertices
but not in the baryon-to-baryon matrix elements (e.g. masses and
weak transition elements). In the case of ground-state intermediate
baryons this leads to the effective {\em subtraction} of the
spin-flavour factors corresponding to diagrams $(b1)$ and $(b2)$.
On the other hand, a closer look at Table 10 and Eq.(7) of
ref.~\cite{KK92} reveals that in the approach of K\"{o}rner these
factors are {\em added}. The net outcome of this difference is
probably most easily seen on the example of the
$\Lambda ^{+}_{c}~\rightarrow ~\Lambda ~\pi ^{+}$ decay.
Namely, it is well known that
the parity conserving amplitude of this decay receives no contributions
from the baryon pole terms in the appropriate symmetry limit \cite
{PTR90}. In refs.~\cite{KKW79,KK92,KS91} the total contribution
from diagrams $(b1)$ and $(b2)$ is, however, nonvanishing. In other
words, in refs.~\cite{KKW79,KK92,KS91} the intermediate baryons are
all assumed to be much heavier than the external ground-state baryons.

One encounters the latter situation e.g. in the parity violating
hyperon decay amplitudes. There, the intermediate $\frac{1}{2}^{-}$
excited baryons are indeed heavier than the external ground-state
$\frac{1}{2}^{+}$    baryons. However, for parity conserving hyperon
decay amplitudes the assumptions of refs.~\cite{KKW79,KK92,KS91}
correspond to neglecting the dominant contribution arising from the
intermediate ground-state baryons (which is singular in the flavour
symmetry limit). Consequently, it is the prescription of the previous
subsection (i.e. Section \ref{pc ampl}) and not that of
refs.~\cite{KKW79,KK92,KS91}
that corresponds to the symmetry structure of the standard pole model
of the parity conserving amplitudes.

The agreement of the symmetry structure of the parity violating
hyperon decay amplitudes as calculated in the quark and pole models
is thus - to some extent - accidental. Namely, had the "excited"
$B^{*}(\frac{1}{2}^{-})$ baryons been degenerate with the ground-state
$B(\frac{1}{2}^{+})$ baryons (but assuming broken $SU(3)$ i.e.
$\Lambda ^{*} = \Lambda = \Sigma ^{*} = \Sigma > N^{*} = N$ ) we
would have ended up with an analogous situation in the parity
violating sector (see also ref.~\cite{Gai88}).

As it is obvious from the above discussion, in general both the
parity violating and the parity conserving amplitudes may contain
pieces with symmetry structure of both the sum and the difference
of spin-flavour factors of diagrams $(b1)$ and $(b2)$. Which of the
two is dominant (if any) depends on the dynamics. Similar considerations
apply of course also to diagram $(d)$. The smallness of its contribution
to the parity conserving amplitudes - as obtained in the fit of
ref.~\cite{KK92} - should perhaps be understood as an   indication
of the dominance of the "difference" structure in diagram $(d)$, in
agreement with
the prescriptions of the standard pole model with intermediate
ground-state baryons. Clearly, the smallness of diagrams $(d)$
obtained in ref.~\cite{KK92} cannot be understood as a complete
phenomenological "proof"
of the dominance of this "difference" structure since in their fit
K\"{o}rner and Kramer used the "sum" structure for diagrams $(b)$.

It is very unfortunate that the highlighted above essential
difference between the (naively applied) arguments of symmetry
and the structure of the standard pole model - although recognized
already in the classical treatise of Marshak, Riazuddin and Ryan
\cite{MRR69} - has been forgotten in various later papers and
books on the subject (see e.g. ref.~\cite{Okun}).

\subsection{The $SU(4)$ Link between the Hyperon and Charmed Baryon
Decays}
\label{SU(4)link}
To calculate the absolute size of the nonleptonic decays of charmed
baryons some authors (e.g. ref.~\cite{PTR90}, for other references
see ref.~\cite{XK92}) use $SU(4)$ symmetry to get the relevant
information from hyperon decays. The way in which $SU(4)$ symmetry
is applied in such approaches is equivalent to the consideration
of symmetry relationships between the baryon-to-baryon matrix elements
of the parity conserving part of the $W$-exchange contribution.
The relevant diagrams are shown in Fig.~2.1.

Diagrams of Fig.~2.1 lead to the well-known $SU(4)$ relation which connects
charmed-baryon and hyperon nonleptonic decays:

\begin{equation}
\label{eq:s1}
<\Sigma ^{+}| H^{p.c.}_{weak} | \Lambda ^{+}_{c} > = \frac{1}{\sqrt{6}}
\cot \theta~_{c} < p | H^{p.c.}_{weak} | \Sigma~^{+} >.
\end{equation}

It has been argued \cite{XK92} that $SU(4)$ symmetry breaking due to the large
mass difference between $c$ and $s$ quarks should lead to a large
mismatch in the baryon wave functions used in the overlap integrals
in Eq.(\ref{eq:s1}). As a result the baryon matrix elements
of the $\Delta C~=~+1$ weak Hamiltonian should be smaller than that
given by Eq.(\ref{eq:s1}). Estimates in the bag model \cite{EK83,Che85,XK92}
yield a correction factor of around 0.5.

Here we point out another reason why these matrix elements should
be smaller than expected on the basis of Eq.(\ref{eq:s1}). Namely,
in the quark model/symmetry approach of refs~\cite{DDH80,Zen89}
there is a large contribution to the
$<~p~|~H^{p.c.}_{weak}~|~\Sigma ^{+}~>$ matrix element that comes
from the "sea-quark"
diagrams (Fig.~2.2 or Fig.~1.c). On the other hand, in charmed
baryon decays the $(c)$-type  diagrams are absent. Consequently,
one has to "subtract" from the experimental value of the
$<~p~|~H^{p.c.}_{weak}~|~\Sigma ^{+}~>$ matrix
element this part of it that is due to diagram~2 in Fig.~2. This
leads to the replacement of formula (\ref{eq:s1}) by:
\begin{equation}
\label{eq:s2}
<~\Sigma ^{+}~|~H^{p.c.}_{weak}~|~\Lambda ^{+}_{c}~> =
 \frac{2}{1 - (\frac{f}{d})_{soft meson}}            \frac{1}{\sqrt{6}}
\cot \theta _{c} <~p~|~H^{p.c.}_{weak}~|~\Sigma ^{+}~>.
\end{equation}
where $(\frac{f}{d})_{soft meson}$ is the ratio
                                              of the invariant $SU(3)$
couplings $f$ and $d$ in the soft meson approximation to the parity
violating amplitudes of nonleptonic hyperon decays (or in the
baryon-to-baryon matrix elements of the parity conserving part of the
$\Delta~S~=~1$ weak Hamiltonian). Estimates of
$(\frac{f}{d})_{soft meson}$ vary.
If one uses the estimate of ref.~\cite{Pha84,Zen92} ($f/d=-1.6$) one gets
a suppression factor of
\begin{equation}
\frac{2}{1-f/d}~\rightarrow~0.77.
\end{equation}
If, on the other hand, one assigns the whole experimentally observed
deviation from $f/d=-1$ in the parity violating amplitudes ($(f/d)~
_{p.v.}~=~-2.5$) to the soft-meson term (and nothing to other possible
terms) one obtains
\begin{equation}
\frac{2}{1-f/d}~\rightarrow~0.56.
\end{equation}
Apparently, "subtraction" of this part of the $f$ coupling that does
not come from the $W$-exchange diagram leads to a very substantial
correction to the naive $SU(4)$ formula (\ref{eq:s1}).

The origin of the deviation of $f/d$ from its naive quark model value
of $-1$ has not been agreed upon yet. We think that the main correction
is due to the sea quark effects discussed in refs~\cite{Don77,Zen92}.
Such effects not only renormalize the $f/d$ ratio but - on account of
the large mass of the charmed quark - might renormalize differently
the $W$-exchange diagrams in the $\Sigma ^{+}~\rightarrow p$ and
$\Lambda ^{+}_{c}~\rightarrow \Sigma ^{+}$ transitions ( a part of the
sea  contribution in the latter
-the $c\bar{c}$ sea- should be negligible). We have checked by explicit
calculation,     however, that in the framework of the hadron-loop
model for the quark sea (ref.~\cite{Zen92,Tor85}), the resulting
renormalization of both transitions are identical precisely when the
$c\bar{c}$ sea is neglected.

The effect discussed above has been taken into account in Section
\ref{pc ampl} where we related the size of the reduced matrix elements
in the hyperon and charmed-baryon parity conserving amplitudes
(Eqs.(\ref{eq:b6}) and (\ref{eq:b7})).

\subsection{Factorization and $SU(2)_{W}$ in Parity Conserving
Amplitudes}
\label{factor-pc}
Evaluation of the diagrams $(a)$ and $(a')$ in Fig.~1 is most widely
performed through the use of the factorization technique. In this
approach one starts with the QCD-corrected effective weak Hamiltonian
which, for the $\Delta C~= \Delta S~=~+1$ processes in question takes
the form:

\begin{equation}
\label{eq:s3}
H_{weak} = \frac{G \cos ^{2} \theta _{c}}{\sqrt{2}}
           (c_{1} O_{1} + c_{2} O_{2} )
\end{equation}

where

\begin{eqnarray}
\label{eq:s4}
O_{1} & = & [\bar{s}\gamma _{\mu}(1-\gamma _{5})c]
        [\bar{u}\gamma _{\mu}(1-\gamma _{5})d] \nonumber \\
O_{2} & = & [\bar{u}\gamma _{\mu}(1-\gamma _{5})c]
        [\bar{s}\gamma _{\mu}(1-\gamma _{5})d]
\end{eqnarray}

The Wilson coefficients $c_{1}$, $c_{2}$ include the short-range QCD effects
and for charm decays they have the values
\begin{eqnarray}
\label{eq:s5}
c_{1} & \approx & 1.3 \nonumber \\
c_{2} & \approx & -0.6
\end{eqnarray}

In accordance with the factorization idea the $(\bar{u}d)$ current in
$O_{1}$ [$(\bar{s}d)$ in $O_{2}$] generates $\pi ^{+}$ or $\rho ^{+}$
[ $\bar{K} ^{0}$ or $\bar{K^{*}}^{0}$]
                         out of hadronic vacuum. (This is the so-called
"new factorization" in which the Fierz-transformed contributions from
Eq.(\ref{eq:s4}) are simply dropped. Such an assumption has now
considerable experimental support \cite{KK92,XK92,CT92}.) Operator
$O_{1}$ corresponds to diagram $(a')$, while
                       $ O_{2}$            to diagram $(a)$ after its
"customization" by Fierz-transformation to the needs of factorization
technique.

Let us consider the factorization contribution to the parity conserving
amplitudes of the $\Lambda ^{+}_{c}~\rightarrow \Lambda \pi ^{+}$ and
$\Lambda ^{+}_{c}~\rightarrow \Lambda \rho ^{+}$ decays. From
Eq.(\ref{eq:s3}) one obtains then

\begin{eqnarray}
\label{eq:s6}
<~\pi ^{+}~\Lambda |~H^{p.c.}_{weak}~|~\Lambda ^{+}_{c}~>_{fact.}
& = & \frac{G \cos ^{2} \theta _{c}}{\sqrt{2}} c_{1}
<\pi ^{+}|~A^{\mu }|~0> <\Lambda |~A_{\mu }|\Lambda ^{+}_{c}>
\nonumber \\
<~\rho ^{+}~\Lambda |~H^{p.c.}_{weak}~|~\Lambda ^{+}_{c}~>_
{fact.} &
= & \frac{G \cos ^{2} \theta _{c}}{\sqrt{2}} c_{1}
<\rho ^{+}|~V^{\mu }|~0> <\Lambda |~V_{\mu }|\Lambda ^{+}_{c}>
\nonumber \\
\phantom{L} & &
\end{eqnarray}

The matrix elements of the currents in Eq.(\ref{eq:s6}) are given by

\begin{eqnarray}
\label{eq:s7}
<~\pi ^{+}~|~A^{\mu }~|~0~> & = & f_{\pi } q^{\mu } \nonumber \\
<~\rho ^{+}~|~V^{\mu }~|~0~> & = & \epsilon ^{\mu } f_{\rho } \nonumber
\\
<~\Lambda ~|~A_{\mu }~|~\Lambda ^{+}_{c}~> & = &
 g^{A}_{\Lambda \Lambda ^{+}_{c}
}(m^{2}_{\pi}) \bar{u} _{\Lambda }\gamma _{\mu}\gamma _{5}
u_{\Lambda ^{+}_{c}} \nonumber \\
<~\Lambda ~|~V_{\mu }~|~\Lambda ^{+}_{c}~> & = & f^{V}_{\Lambda \Lambda
^{+}_{c}
} (m^{2}_{\rho}) \bar{u}_{\Lambda} \gamma _{\mu} u_{\Lambda ^{+}_{c}}
\end{eqnarray}

where $f_{\pi} = 0.13 GeV$, $f_{\rho} = 0.17 GeV^{2}$ and $g^{A}$,
$f^{V}$ are axial-vector and vector formfactors.

Let us now see if the factorization prescription is consistent with
the $SU(2)_{W}$ symmetry between the $\pi ^{+}$ and $\rho ^{+}$ couplings
as employed in the previous Section.
Application of the requirement of $SU(2)_{W}$ symmetry to the couplings
of Eq.(\ref{eq:s6}) leads to the following condition:

\begin{equation}
\label{eq:s8}
f_{\rho} = f_{\pi} (m_{\Lambda}+m_{\Lambda ^{+}_{c}})
\frac{g^{A}_{\Lambda \Lambda ^{+}_{c}}(m^{2}_{\pi})}
    {f^{V}_{\Lambda \Lambda ^{+}_{c}}(m^{2}_{\rho})}
\end{equation}

The ratio of $g^{A}/f^{V}$ is 1 in the simplest approach. If the bag
model calculations of these formfactors are employed (ref.~\cite
{AviAo}) one obtains instead (with $g^{A}_{\Lambda \Lambda ^{+}_{c}} =
0.50$, $f^{V}_{\Lambda \Lambda ^{+}_{c}} = 0.46$)

\begin{eqnarray}
\label{eq:s9}
g^{A}_{\Lambda \Lambda^{+}_{c}}(m^{2}_{\pi}) & = &
g^{A}_{\Lambda \Lambda^{+}_{c}}(1-\frac{m^{2}_{\pi}}{m^{2}_{A}})^{-2}
 \approx ~0.50~~~~~~~~(m_{A}=2.54 GeV) \nonumber \\
f^{V}_{\Lambda \Lambda^{+}_{c}}(m^{2}_{\rho}) & = &
f^{V}_{\Lambda \Lambda^{+}_{c}}(1-\frac{m^{2}_{\rho}}{m^{2}_{*}})^{-2}
 \approx ~0.61~~~~~~~~(m_{*}=2.11 GeV)
\end{eqnarray}

and the relevant ratio of axial and vector formfactors becomes
smaller:

\begin{equation}
\frac{g^{A}_{\Lambda \Lambda^{+}_{c}}(m^{2}_{\pi})}
{f^{V}_{\Lambda \Lambda^{+}_{c}}(m^{2}_{\rho})} = 0.82
\end{equation}

Using the above value of $g^{A}/f^{V}$ equation~(\ref{eq:s8}) then reads:
\begin{equation}
\label{eq:s10}
0.17 Gev^{2} = 0.36 GeV^{2}
\end{equation}

There is therefore a factor of 2 discrepancy (2.5 if $g^{A}/f^{V}=1$ is
used) between the $SU(2)_{W}$ symmetry predictions and the standard
factorization technique. Similar discrepancy exists between the
$SU(2)_{W}$ and factorization predictions for the $\bar{K}^{o}$ and
$\bar{K^{*}}^{o}$ production amplitudes of diagram $(a)$. One has to keep
in mind, however, that - in principle - the factorization amplitude
constitutes but a single contribution to the meson-pole terms \cite
{DGH86}. Unfortunately, direct theoretical estimates of these
contributions do not seem to be reliable \cite{DGH86}. If one believes
in the accuracy of the $SU(2)_{W}$ symmetry predictions, the
discrepancy of Eq.(\ref{eq:s10}) shows that the factorization technique
may be trusted here to within a factor of 2 only. Such accuracy is
insufficient for making reliable predictions. On the other hand, if the
contributions from the $f_{2}\bar{u}_{f}\sigma _{\mu \nu}q^{\nu}u_{i}$
and $g_{2}\bar{u}_{f}\sigma _{\mu \nu} \gamma _{5} q^{\nu}u_{i}$ terms
to the current matrix elements are considered (as in ref.~\cite{CT92})
the disagreement in question is much milder ( $\approx$ 30\% ).

\subsection{Factorization and Sextet Dominance}
\label{factor sextet}
The relative size of the factorization contribution to the nonstrange
($\pi ^{+}, \rho ^{+}$) and strange ($\bar{K}^{o}, \bar{K^{*}}^{o}$)
meson emission is fixed by (\ref{eq:s5}) and $SU(3)$ symmetry-breaking
factors like $f_{K}/f_{\pi}$ as well as by the $q^{2}$-dependence of the
formfactors $g^{A}$ and $f^{V}$. Calculations along these lines are
straightforward (e.g. see ref.~\cite{CT92}). To relate such calculations
to the parametrization of this paper we express below the results of
refs.~\cite{XK92,CT92} in terms of our reduced matrix elements.

For the parity conserving amplitudes the estimates of the
factorization amplitudes of Cheng and Tseng \cite{CT92} correspond
to the following values of the $M,M'$ parameters of Section~\ref{basic
model} (in units of $10^{-7}$):

{\it (a)} for the pseudoscalar mesons
\begin{eqnarray}
\label{eq:s11}
M & \approx & 75 \nonumber \\
M'& \approx & -120
\end{eqnarray}

{\it (b)} for the transverse vector mesons
\begin{eqnarray}
\label{eq:s12}
M & \approx & 61 \nonumber \\
M'& \approx & -88
\end{eqnarray}

One observes that $(M'/M)_{CT} \approx -1.5$, not very far from the
sextet-dominance relation $M'/M=-1$.

For the longitudinal vector mesons sextet dominance requires
similarly $m'/m=-1$, while the quark model relates the reduced
matrix elements $m$, $M$ by

\begin{equation}
\label{eq:s13}
m = -M
\end{equation}

For the parity conserving amplitudes the calculations of ref.~\cite
{CT92} correspond to (in units of $10^{-7}$)
\begin{eqnarray}
\label{eq:s14}
g_{CT}  & \approx & 5 \nonumber \\
g'_{CT} & \approx & -6
\end{eqnarray}

while those of ref.~\cite{XK92} yield
\begin{eqnarray}
\label{eq:s15}
g_{XK} & \approx & 3.4 \pm 1 \nonumber \\
g'_{XK} & \approx & -6.5
\end{eqnarray}

In the vector meson sector results of ref.~\cite{CT92} are
translated into our scheme as
\begin{eqnarray}
\label{eq:s16}
\tilde{a}_{CT} & = & 9.6 \nonumber \\
\tilde{a'}_{CT} & = & -14.3
\end{eqnarray}

Again, the ratios $\tilde{a'}/\tilde{a}$ or $g'/g$ are around -1.5,
not very far from the sextet dominance value of -1. The estimates
of Eq.(\ref{eq:s16}) correspond to

\begin{eqnarray}
\label{eq:s17}
a_{CT} & = & +2.2 \nonumber \\
a'_{CT} & = & -3.2
\end{eqnarray}

These numbers should be compared with an estimate of Desplanques,
Donoghue and Holstein \cite{DDH80}
\begin{equation}
\label{eq:s18}
a'_{DDH} \approx -3.0
\end{equation}

and with the result of the fit to the weak radiative hyperon decays
\cite{Zen91}
\begin{equation}
\label{eq:s19}
a_{Z} = +3.8
\end{equation}

In view of the inherent uncertainties of the factorization technique all
these estimates suggest that the sextet-dominance assumption may
be a good approximation for the "factorization" amplitudes of the
nonleptonic decays of charmed baryons. Similar view has been
expressed by Savage and Springer \cite{SS90}.

\section{Fits and Conclusions}
\label{fits}
In the preceding Section it was pointed out that in the symmetry
approach of K\"{o}rner and collaborators the symmetry structure of
the parity conserving amplitudes does not correspond to the symmetry
structure of the standard pole model. This fact plus the appearance
of various uncertainties in the reduced matrix elements under consideration
(as also discussed in the last Section) means that the fits in the
symmetry-based approach should be done anew.
In the following we will present such a fit. We stress very strongly,
however, that - on account of many simplifications involved - the fit
should not be considered overly seriously. Rather it should be regarded
as purporting the thesis that the present data on charmed baryon decays
can be well accommodated in the symmetry-based approach. The fairly
limited set of data now available does not warrant a detailed
consideration of various symmetry breaking effects. It is only when
more data are gathered that the phenomenological determination and
discussion of such effects will become possible within the
generic framework of this paper. Since at present there
are only a few experimental numbers to be fitted we must reduce the
number of free parameters of the fit if it is to be meaningful.
To this end we make the following simplifying assumptions:

(1) we assume that the connection between the longitudinal and
transverse vector meson emission is that given by the quark model, i.e.

\begin{eqnarray}
\label{eq:f0}
\tilde{b}' & = & \tilde{b} \nonumber \\
B' &  = & B \nonumber \\
m & = & -M
\end{eqnarray}

Our fit is based on four fairly accurate data points characterising the
decays with the emission of pseudoscalar mesons and on
the not so well determined branching ratio for the
$\Lambda ^{+}_{c}~\rightarrow p~\phi$ process. Consequently, the
above assumption does not affect the predictions of the fit
for the decays with the emission of pseudoscalar mesons.

(2) we assume that the sextet dominance rule holds for the factorization
contributions to the parity conserving amplitudes with both pseudoscalar
and vector meson emission, i.e.:
\begin{eqnarray}
\label{eq:f1}
M' & = & -M \nonumber \\
m' & = & -m (=M)
\end{eqnarray}

as well as for the factorization contributions to the parity violating
amplitudes with pseudoscalar meson emission

\begin{equation}
\label{eq:f2}
g' = -g
\end{equation}

For the factorization pieces in the parity violating amplitudes with
(transverse) vector meson emission we use the values extracted from
hyperon decays (Eqs.(\ref{eq:s18},\ref{eq:s19})).
No significant change in
the quality of the fit to the $\Lambda ^{+}_{c}~\rightarrow p~\phi$
is observed if one accepts sextet dominance for these amplitudes
with $\tilde{a}~= -\tilde{a}'~\approx 13$. To further diminish the
number of free parameters we use a single value of g in the range
suggested in Eqs.(\ref{eq:s14}) and (\ref{eq:s15}):
\begin{equation}
\label{eq:f3}
g = 4.5
\end{equation}
The above assumption of sextet dominance seems acceptable in view of
the inherent uncertainty of the factorization estimates (\ref{eq:s11})
,(\ref{eq:s12}). Furthermore, it reduces the number of free parameters
significantly.

(3) The values of parameters corresponding to the $W$-exchange diagrams
$(b)$ should be taken from hyperon decays (i.e. $b = -5.0$, $B = -97.5$,
Eqs.(\ref{eq:b2a}) and (\ref{eq:b8})). However, as discussed by Xu and
Kamal \cite{XK92}, one expects a mismatch in the baryon wave functions
of charmed and noncharmed baryons due to the large mass of the charmed
quark. Thus, one expects the $W$-exchange contributions to be smaller
than the simple estimates of Eqs.(\ref{eq:b2a}) and (\ref{eq:b8}).
We take this into account by introducing an overlap parameter r
such that the reduced
matrix elements $b,B$ are replaced in our formulas by
\begin{eqnarray}
\label{eq:f4}
b & \rightarrow & r b \nonumber \\
B & \rightarrow & r B
\end{eqnarray}

In the following we fit the absolute branching ratios given by Particle
Data Group \cite{PDG92}. One has to remember, though, that these are
measured relative to the $\Lambda _{c}^{+}~\rightarrow p~K^{-}~\pi ^{+}$
\cite{KS91}. Thus, the fitted value of $r$ does not correspond to the
overlap suppression factor alone - it takes care of the uncertainty in
the absolute size of the experimental branching ratios as well.

In summary, we have two parameters: $M$ and $r$ and five experimental
data points to be fitted. These are the branching ratios of
$\Lambda _{c}^{+}~\rightarrow \Sigma ^{0}~\pi ^{+}, \Lambda \pi ^{+},
p\bar{K^{0}}, p\phi$ and the asymmetry of the
$\Lambda _{c}^{+}~\rightarrow \Lambda \pi ^{+} $ process.
The fit achieves $\chi ^{2}~\approx~1.2$ with three degrees of freedom.
Apparently, the data are
not restrictive enough as yet. Results of the fit are presented in
Table 3 and compared with other papers in Table 4. The fitted values
of parameters are:
\begin{eqnarray}
\label{eq:f5}
r_{fit} & = & 0.63 \nonumber \\
M_{fit} & = & +45.
\end{eqnarray}

As expected $r$ is smaller than $1$. The fitted value of the reduced
matrix element $M$ is about half of that predicted in the factorization
approach (c.f. Eqs.(\ref{eq:s11}) and (\ref{eq:s12})). One has to remember,
however, that

(1) as it was argued in Section~\ref{factor-pc} factorization may be
trusted to within a factor of 1.5 or 2.

(2) the uncertainty in the absolute size of the branching ratios has
not been taken into account here (as it was the case for the reduced
matrix elements $\tilde{b}$ and $B$ since $M$ is a free parameter,
anyway.

Although the presented fit is obtained under several simplifying
assumptions, it suggests that factorization amplitudes are not
as big as one might expect. That factorization prescription seems to
give too large contributions has been already noticed by Ebert and
Kallies \cite{EK83}. Furthermore, the fit indicates that nonvanishing
contributions from the W-exchange diagram $(b)$ are needed. Their
presence thwarts all attempts to describe nonleptonic decays of
charmed baryons with the help of the factorization contribution
alone. This conclusion was stressed in refs.~\cite{KK92,KS91} as well.

In summary, we have shown that the parity conserving amplitudes in
the symmetry approach of refs.~\cite{KKW79,KK92,KS91} do not possess
the symmetries of the standard pole model with ground-state intermediate
baryons. The proper symmetry structure of these amplitudes that does
correspond to this standard assumption of the pole model has been given.
In addition, a couple
of uncertainties inherent in present approaches to nonleptonic decays of
charmed baryons have
been identified and discussed. Finally, a fit to the existing data has
been carried out.

We would like to stress once again that the fit itself should not be
taken overly seriously. There are many unanswered questions concerning
the validity of the adopted assumptions. For example, one may worry about
the contributions to the parity-conserving amplitudes from intermediate
baryons other than the ground-state ones, such as members of the
radially excited $(56,\frac{1}{2}^{+*})$ multiplet \cite{Tur91}.
Another questionable assumption is that of the $SU(4)$ current
algebra used in the description of parity violating amplitudes: it is
only in the limit of exact $SU(4)$ that current algebra and the
standard $\frac{1}{2}^{-}$ pole model become equivalent. Although
further theoretical studies of various such symmetry breaking
effects in the general framework adopted in this paper are clearly
needed, we believe that it will be the experiment that will guide
us in our attempts to understand theoretically the nonleptonic
decays of charmed baryons.

%
\renewcommand{\baselinestretch}{2}
\small
\normalsize
\newpage
Table 1. Weak amplitudes for $B_{c} \rightarrow BP$ decays.
\\
\\
\begin{tabular}{||l||c ||c| c|c||}
\hline
process & $A_{P}$   & diag. $(b1)$ & diag. $(b2)$  &  $B_{P}$            \\
\hline
$\Xi ^{+}_{c} \rightarrow \Xi ^{0} \pi ^{+}$ & $-\frac{1}{2\sqrt{6}}\tilde{b}
+ \sqrt{\frac{3}{2}} g'$ & $0$ & $-\frac{1}{2\sqrt{6}}$ & $\frac{1}{2\sqrt{6}
}
B-\frac{1}{2\sqrt{6}} M'$ \\
$\Xi ^{+}_{c} \rightarrow \Sigma ^{+} \overline{K}^{0}$ & $-\frac{1}{2\sqrt{6}}
\tilde{b}-\sqrt{\frac{3}{2}} g$ & $0$ & $-\frac{1}{2\sqrt{6}}$ & $\frac{1}{2
\sqrt{6}}B+\frac{1}{2\sqrt{6}}M$ \\ \hline
$\Xi ^{0}_{c} \rightarrow \Xi ^{0} \pi ^{0}$ & $-\frac{1}{2\sqrt{3}} \tilde{b}$
& $\frac{1}{12\sqrt{3}}$ & $-\frac{1}{4\sqrt{3}}$ & $\frac{1}{3\sqrt{3}} B$ \\
$\Xi ^{0}_{c} \rightarrow \Xi ^{0} \eta _{8}$ & $\frac{1}{6}\tilde{b}$ &
$\frac{1}{12}$ & $\frac{1}{12}$ & $0$ \\
$\Xi ^{0}_{c} \rightarrow \Xi ^{0} \eta _{1}$ & $-\frac{1}{6\sqrt{2}}
\tilde{b}$ & $0$ & $\frac{1}{6\sqrt{2}}$ & $-\frac{1}{6\sqrt{2}}B$ \\
$\Xi ^{0}_{c} \rightarrow \Xi ^{-} \pi ^{+}$ & $\frac{1}{2\sqrt{6}}\tilde{b}
+\sqrt{\frac{3}{2}}g'$ & $-\frac{1}{6\sqrt{6}}$ & $0$ & $-\frac{1}{6\sqrt{6}}
B-\frac{1}{2\sqrt{6}}M'$ \\
$\Xi ^{0}_{c} \rightarrow \Sigma ^{0} \overline{K}^{0}$ & $-\frac{1}{4
\sqrt{3}}\tilde{b}-\frac{\sqrt{3}}{2}g$ & $-\frac{1}{6\sqrt{3}}$ &
$-\frac{1}{4\sqrt{3}}$ & $\frac{1}{12\sqrt{3}}B+\frac{1}{4\sqrt{3}}M$ \\
$\Xi ^{0}_{c} \rightarrow \Lambda \overline{K}^{0}$ & $-\frac{1}{4}\tilde{b}
+\frac{1}{2}g$ & $0$ & $-\frac{1}{12}$ & $\frac{1}{12}B-\frac{1}{12}M$ \\
$\Xi ^{0}_{c} \rightarrow \Sigma ^{+} K^{-}$ & $0$ &
$\frac{1}{3\sqrt{6}}$ & $0$ & $\frac{1}{3\sqrt{6}}B$ \\ \hline
$\Lambda ^{+}_{c} \rightarrow \Sigma ^{+} \pi ^{0}$ & $-\frac{1}{2\sqrt{3}}
\tilde{b}$ & $-\frac{1}{12\sqrt{3}}$ & $-\frac{1}{4\sqrt{3}}$ &
$\frac{1}{6\sqrt{3}}B$ \\
$\Lambda ^{+}_{c} \rightarrow \Sigma ^{+} \eta _{8}$ & $\frac{1}{6}\tilde{b}$
& $-\frac{1}{12}$ & $\frac{1}{12}$ & $-\frac{1}{6}B$ \\
$\Lambda ^{+}_{c} \rightarrow \Sigma ^{+} \eta _{1}$ & $-\frac{1}{6\sqrt{2}}
\tilde{b}$ & $0$ & $\frac{1}{6\sqrt{2}}$ & $-\frac{1}{6\sqrt{2}}B$ \\
$\Lambda ^{+}_{c} \rightarrow \Sigma ^{0} \pi ^{+}$ & $\frac{1}{2\sqrt{3}}
\tilde{b}$ & $\frac{1}{12\sqrt{3}}$ & $\frac{1}{4\sqrt{3}}$ &
$-\frac{1}{6\sqrt{3}}B$ \\
$\Lambda ^{+}_{c} \rightarrow \Lambda \pi ^{+}$ & $-g'$ & $\frac{1}{12}$ &
$\frac{1}{12}$ & $\frac{1}{6}M'$ \\
$\Lambda ^{+}_{c} \rightarrow \Xi ^{0} K^{+}$ & $0$ & $\frac{1}{3\sqrt{6}}$
& $0$ & $\frac{1}{3\sqrt{6}}B$ \\
$\Lambda ^{+}_{c} \rightarrow p \overline{K}^{0}$ & $\frac{1}{2\sqrt{6}}
\tilde{b}-\sqrt{\frac{3}{2}}g$ & $-\frac{1}{6\sqrt{6}}$ & $0$ &
$-\frac{1}{6\sqrt{6}}B+\frac{1}{2\sqrt{6}}M$ \\ \hline
\end{tabular}
\newpage
Table 2. Weak amplitudes for $B_{c} \rightarrow BV$ decays.
\\
\\
\begin{tabular}{||l||c|c|c|c|}
\hline
process & $A_{V_{\perp}}$    & $B_{V_{\perp}}$     & $A_{V_{\parallel}}$ &
$B_{V_{\parallel}}$ \\
\hline
$\Xi ^{+}_{c} \rightarrow \Xi ^{0} \rho ^{+}$ & $-\frac{1}{2\sqrt{3}}\tilde{b}
+\frac{2}{\sqrt{3}}\tilde{a}'$ & $-\frac{1}{2\sqrt{3}}B+\frac{1}{2\sqrt{3}}M'$
& $\frac{1}{2\sqrt{6}}\tilde{b}'$ & $-\frac{1}{2\sqrt{6}}B'+\frac{1}{2\sqrt{6}}
m'$ \\
$\Xi ^{+}_{c} \rightarrow \Sigma ^{+} \overline{K}^{*0}$ & $-\frac{1}{2
\sqrt{3}}\tilde{b}-\frac{2}{\sqrt{3}}\tilde{a}$ & $-\frac{1}{2\sqrt{3}}B-
\frac{1}{2\sqrt{3}}M$ & $\frac{1}{2\sqrt{6}}\tilde{b}'$ & $-\frac{1}{2
\sqrt{6}}B'-\frac{1}{2\sqrt{6}}m$ \\
\hline
$\Xi ^{0}_{c} \rightarrow \Xi ^{0} \rho ^{0}$ & $-\frac{1}{3\sqrt{6}}
\tilde{b}$ &
$-\frac{2}{3\sqrt{6}}B$ & $\frac{1}{3\sqrt{3}}\tilde{b}'$ & $-\frac{1}{2
\sqrt{3}}B'$ \\
$\Xi ^{0}_{c} \rightarrow \Xi ^{0} \omega$ & $+\frac{2}{3\sqrt{6}}\tilde{b}$ &
$+\frac{1}{3\sqrt{6}}B$ & $-\frac{1}{6\sqrt{3}}\tilde{b}'$ & $0$ \\
$\Xi ^{0}_{c} \rightarrow \Xi ^{0} \phi$ & $-\frac{1}{6\sqrt{3}}\tilde{b}$ &
$+\frac{1}{6\sqrt{3}}B$ & $-\frac{1}{6\sqrt{6}}\tilde{b}'$ &
$-\frac{1}{2\sqrt{6}}B'$ \\
$\Xi ^{0}_{c} \rightarrow \Xi ^{-} \rho ^{+}$ & $-\frac{1}{6\sqrt{3}}\tilde{b}+
\frac{2}{\sqrt{3}}\tilde{a}'$ & $\frac{1}{6\sqrt{3}}B+\frac{1}{2\sqrt{3}}M'$ &
$-\frac{1}{6\sqrt{6}}\tilde{b}'$ &
$\frac{1}{2\sqrt{6}}B'+\frac{1}{2\sqrt{6}}m'$
\\
$\Xi ^{0}_{c} \rightarrow \Sigma ^{0} \overline{K}^{*0}$ &
$-\frac{5}{6\sqrt{6}}
\tilde{b}-\sqrt{\frac{2}{3}}\tilde{a}$ & $-\frac{1}{6\sqrt{6}}B-\frac{1}{2
\sqrt{6}}M$ & $\frac{1}{12\sqrt{3}}\tilde{b}'$ & $\frac{1}{4\sqrt{3}}B'-
\frac{1}{4\sqrt{3}}m$ \\
$\Xi ^{0}_{c} \rightarrow \Lambda \overline{K}^{*0}$ & $-\frac{1}{6\sqrt{2}}
\tilde{b}+\frac{2}{3\sqrt{2}}\tilde{a}$ & $-\frac{1}{6\sqrt{2}}B+
\frac{1}{6\sqrt{2}}M$ & $\frac{1}{12}\tilde{b}'$ & $-\frac{1}{4}B'+\frac{1}{12}
m$ \\
$\Xi ^{0}_{c} \rightarrow \Sigma ^{+} \overline{K}^{*-}$ & $\frac{1}{3\sqrt{3}}
\tilde{b}$ & $-\frac{1}{3\sqrt{3}}B$ & $\frac{1}{3\sqrt{6}}\tilde{b}'$ & $0$ \\
\hline
$\Lambda ^{+}_{c} \rightarrow \Sigma ^{+} \rho ^{0}$ & $-\frac{2}{3\sqrt{6}}
\tilde{b}$ & $-\frac{1}{3\sqrt{6}}B$ & $\frac{1}{6\sqrt{3}}\tilde{b}'$ &
$-\frac{1}{2\sqrt{3}}B'$ \\
$\Lambda ^{+}_{c} \rightarrow \Sigma ^{+} \omega$ & $\frac{1}{3\sqrt{6}}
\tilde{b}$ & $\frac{2}{3\sqrt{6}}B$ & $-\frac{1}{3\sqrt{3}}\tilde{b}'$ & $0$ \\
$\Lambda ^{+}_{c} \rightarrow \Sigma ^{+} \phi$ &
$\frac{1}{6\sqrt{3}}\tilde{b}$
& $-\frac{1}{6\sqrt{3}}B$ & $\frac{1}{6\sqrt{6}}\tilde{b}'$ & $-\frac{1}
{2\sqrt{6}}B'$ \\
$\Lambda ^{+}_{c} \rightarrow \Sigma ^{0} \rho ^{+}$ & $\frac{2}{3\sqrt{6}}
\tilde{b}$ & $\frac{1}{3\sqrt{6}}B$ & $-\frac{1}{6\sqrt{3}}\tilde{b}'$ &
$\frac{1}{2\sqrt{3}}B'$ \\
$\Lambda ^{+}_{c} \rightarrow \Lambda \rho ^{+}$ & $\frac{1}{3\sqrt{2}}
\tilde{b}-\frac{4}{3\sqrt{2}}\tilde{a}'$ & $-\frac{1}{3\sqrt{2}}M'$ & $0$ &
$-\frac{1}{6}m'$ \\
$\Lambda ^{+}_{c} \rightarrow \Xi ^{0} K^{*+}$ & $\frac{1}{3\sqrt{3}}
\tilde{b}$ &
$-\frac{1}{3\sqrt{3}}B$ & $\frac{1}{3\sqrt{6}}\tilde{b}'$ & $0$ \\
$\Lambda ^{+}_{c} \rightarrow p \overline{K}^{*0}$ & $-\frac{1}{6\sqrt{3}}
\tilde{b}-\frac{2}{\sqrt{3}}\tilde{a}$ & $\frac{1}{6\sqrt{3}}B-\frac{1}
{2\sqrt{3}}M$ & $-\frac{1}{6\sqrt{6}}\tilde{b}'$ &
$\frac{1}{2\sqrt{6}}B'-
\frac{1}{2\sqrt{6}}m$\\
\hline
$\Lambda ^{+}_{c} \rightarrow p \phi$ & $\frac{2}{\sqrt{3}}\tilde{a}
\tan\theta _{c}$ & $\frac{1}{2 \sqrt{3}} M \tan \theta _{c}$ &
$0$ & $\frac{1}{2 \sqrt{6}} m \tan \theta _{c}$\\
\hline
\end{tabular}
\newpage
Table 3. Fit to branching ratios and asymmetries
\\
\\
\begin{tabular}{||l|c|c||}
\hline
$ \Lambda ^{+}_{c} \rightarrow $ & BR(\% ) & asymmetry \\
\hline
$ \Sigma ^{+} \pi ^{0}$  &   0.43   & -0.76 \\
$ \Sigma ^{+} \eta $     &   0.25   & -0.91 \\
$ \Sigma ^{+} \eta '$    &   0.05   & +0.72 \\
$ \Sigma ^{0} \pi ^{+}$  &   0.43   & -0.76 \\
$ \Lambda \pi ^{+}$      &   0.59   & -0.86 \\
$ \Xi ^{0} K^{+}$        &   0.07   & 0.00  \\
$ p \bar{K}^{0}$         &   1.90   & -0.90 \\
\hline
$ \Sigma ^{+} \rho ^{0}$ &   0.53   & +0.10 \\
$ \Sigma ^{+} \omega $   &   0.36   & +0.57 \\
$ \Sigma ^{+} \phi $     &   0.04   & -0.87 \\
$ \Sigma ^{0} \rho ^{+}$ &   0.53   & +0.10 \\
$ \Lambda \rho ^{+}$     &   0.51   & +0.79 \\
$ \Xi ^{0} K^{*+}$       &   0.09   & -0.54 \\
$ p \bar{K}^{*0}$        &   2.27   & +0.83 \\
\hline
\hline
$ p \phi $               &   0.10   & +0.54 \\
\hline
\end{tabular}
\hfill
\begin{tabular}{||l|c|c||}
\hline
$ \Xi ^{0}_{c} \rightarrow $ & BR(\% ) & asymmetry \\
\hline
$ \Xi ^{0} \pi ^{0} $      &   0.29   & -0.99  \\
$ \Xi ^{0} \eta $          &   0.04   & -0.32  \\
$ \Xi ^{0} \eta ' $        &   0.03   & +0.90  \\
$ \Xi ^{-} \pi ^{+} $      &   0.88   & -0.78  \\
$ \Sigma ^{0} \bar{K}^{0}$ &   0.05   & -0.89  \\
$ \Lambda \bar{K}^{0} $    &   0.40   & -0.84  \\
$ \Sigma ^{+} K^{-}  $     &   0.07   &  0.00  \\
\hline
$ \Xi ^{0} \rho ^{0} $     &   0.31   & -0.17  \\
$ \Xi ^{0} \omega $        &   0.14   & +0.73  \\
$ \Xi ^{0} \phi $          &   0.03   & +0.17  \\
$ \Xi ^{-} \rho ^{+} $     &   0.64   & +0.80  \\
$ \Sigma ^{0} \bar{K}^{*0}$&   0.12   & +0.62  \\
$ \Lambda \bar{K}^{*0} $   &   0.49   & +0.58  \\
$ \Sigma ^{+} K^{*-} $     &   0.14   & -0.81  \\
\hline
\end{tabular}
\\
\vfill
\newpage
Table 3 ({\em cont.})
\\
\\
\begin{tabular}{||l|c|c||}
\hline
$ \Xi ^{+}_{c} \rightarrow $ & BR(\% ) & asymmetry \\
\hline
$ \Xi ^{0} \pi ^{+} $      &   0.31   & +0.65  \\
$ \Sigma ^{+} \bar{K}^{0} $&   0.28   & +0.68  \\
$ \Xi ^{0} \rho ^{+} $     &   1.72   & -0.61  \\
$ \Sigma ^{+} \bar{K}^{*0}$&   2.63   & -0.48  \\
\hline
\end{tabular}

\newpage
Table 4. Comparison of model predictions for selected decays.
\\
\\
\begin{tabular}{|l|cc|cc|cc|cc|}
\hline
 & \multicolumn{2}{c|}{this work} & \multicolumn{2}{c|}{ref. \cite{KK92}} &
\multicolumn{2}{c|}{ref.
\cite{CT92}}
& \multicolumn{2}{c|}{experiment}
\\
$ \Lambda ^{+}_{c} \rightarrow $ & BR & asym & BR & asym & BR & asym &
BR & asym \\
\hline
$ \Sigma ^{+} \pi ^{0} $ & 0.43 & -0.76 & 0.31 & +0.71 & 0.72 & +0.83 & & \\
$ \Sigma ^{+} \eta $     & 0.25 & -0.91 & 0.15 & +0.33 & & & & \\
$ \Sigma ^{+} \eta ' $   & 0.05 & +0.72 & 1.22 & -0.45 & & & & \\
$ \Sigma ^{0} \pi ^{+} $ & 0.43 & -0.76 & 0.31 & +0.70 & 0.72 & +0.83 &
0.55 $\pm $ 0.26 & \\
$ \Lambda \pi ^{+} $     & 0.59 & -0.86 & 0.71 & -0.70 & 0.87 & -0.96 &
0.58 $\pm $ 0.16 & -1.03 $\pm $ 0.29 \\
$ \Xi ^{0} K^{+} $       & 0.07 &  0.00 & 0.25 &  0.00 & & & & \\
$ p \bar{K}^{0} $        & 1.90 & -0.90 & 2.01 & -1.00 & 1.20 & -0.49 &
1.60 $\pm $ 0.40 & \\
\hline
$ \Sigma ^{+} \rho ^{0}$ & 0.53 & +0.10 & 3.0  &     & O(0.1) & +0.10 & & \\
$ \Sigma ^{+} \omega $   & 0.35 & +0.57 & 3.8 &      &   & & &  \\
$ \Lambda \rho ^{+} $    & 0.51 & +0.79 & 18.2 & & 2.3-2.6 & -0.2  & & \\
$ \Xi ^{0} \bar{K}^{*+}$ & 0.09 & -0.54 & 0.11 & & & & &  \\
$ p \bar{K}^{*0} $   & 2.27 & +0.83 & 2.9 & & 1.8-3.3 & -0.1  & & \\
\hline
$ p \phi $      & 0.10 & +0.54 & 0.20 & & 0.19 & & 0.13 $\pm $ 0.9 & \\
\hline
\end{tabular}
\newpage
Table 4. ({\em cont})
\\
\\
\begin{tabular}{|l|cc|cc|}
\hline
 & \multicolumn{2}{c|}{this work} & \multicolumn{2}{c|}{ref. \cite{KK92}} \\
$ \Xi ^{+}_{c} \rightarrow $ & BR & asym & BR & asym \\
\hline
$ \Xi ^{0} \pi ^{+} $       & 0.30 & +0.65 & 2.4 & -0.78 \\
$ \Sigma ^{+} \bar{K}^{0} $ & 0.28 & +0.69 & 4.4 & -1.0  \\
$ \Xi ^{0} \rho ^{+} $      & 1.72 & -0.61 & 65.0 &  \\
$ \Sigma ^{+} \bar{K}^{*0}$ & 2.63 & -0.48 & 1.6 &   \\
\hline
\end{tabular}
\newpage
%
%
\setlength{\unitlength}{0.7pt}
\begin{picture}(470,750)
\put(10,40){
\begin{picture}(450,700)
\put(0,540){
\begin{picture}(200,160)
\put(100,15){\makebox(0,0){(a)}}
\put(30,90){\line(1,0){25}}
\put(85,90){\vector(-1,0){30}}
\put(115,90){\line(1,0){30}}
\put(170,90){\vector(-1,0){25}}
\put(85,150){\vector(0,-1){30}}
\put(85,120){\line(0,-1){30}}
\put(115,90){\vector(0,1){30}}
\put(115,120){\line(0,1){30}}
\put(170,65){\vector(-1,0){70}}
\put(100,65){\line(-1,0){70}}
\put(170,40){\vector(-1,0){70}}
\put(100,40){\line(-1,0){70}}
\multiput(85,90)(5,0){6}{\line(1,0){3}}
\end{picture}}
\put(250,540){
\begin{picture}(200,160)
\put(100,15){\makebox(0,0){(a')}}
\put(170,65){\vector(-1,0){70}}
\put(100,65){\line(-1,0){70}}
\put(170,40){\vector(-1,0){70}}
\put(100,40){\line(-1,0){70}}
\put(85,150){\vector(0,-1){20}}
\put(115,150){\line(0,-1){20}}
\put(115,130){\vector(0,1){0}}
\put(100,130){\oval(30,30)[b]}
\put(170,90){\vector(-1,0){25}}
\put(145,90){\vector(-1,0){90}}
\put(55,90){\line(-1,0){25}}
\multiput(100,90)(0,5){5}{\line(0,1){3}}
\end{picture}}
\put(125,0){
\begin{picture}(200,160)
\put(100,15){\makebox(0,0){(d)}}
\put(30,90){\line(1,0){25}}
\put(85,90){\vector(-1,0){30}}
\put(115,90){\line(1,0){30}}
\put(170,90){\vector(-1,0){25}}
\put(85,150){\vector(0,-1){30}}
\put(85,120){\line(0,-1){30}}
\put(115,90){\vector(0,1){30}}
\put(115,120){\line(0,1){30}}
\put(170,65){\vector(-1,0){70}}
\put(100,65){\line(-1,0){70}}
\put(170,40){\vector(-1,0){70}}
\put(100,40){\line(-1,0){70}}
\multiput(100,40)(0,5){5}{\line(0,1){3}}
\end{picture}}
\put(0,360){
\begin{picture}(200,160)
\put(100,15){\makebox(0,0){(b1)}}
\put(170,90){\vector(-1,0){25}}
\put(145,90){\line(-1,0){50}}
\multiput(130,65)(0,5){5}{\line(0,1){3}}
\put(65,90){\vector(-1,0){20}}
\put(45,90){\line(-1,0){15}}
\put(65,150){\vector(0,-1){30}}
\put(65,120){\line(0,-1){30}}
\put(95,90){\vector(0,1){30}}
\put(95,120){\line(0,1){30}}
\put(170,65){\vector(-1,0){90}}
\put(80,65){\line(-1,0){50}}
\put(170,40){\vector(-1,0){90}}
\put(80,40){\line(-1,0){50}}
\end{picture}}
\put(0,180){
\begin{picture}(200,160)
\put(100,15){\makebox(0,0){(c1)}}
\put(170,90){\vector(-1,0){15}}
\put(155,90){\line(-1,0){10}}
\put(115,90){\line(-1,0){20}}
\multiput(115,90)(5,0){6}{\line(1,0){3}}
\put(130,90){\oval(30,30)[b]}
\put(130,75){\vector(-1,0){0}}
\put(65,90){\vector(-1,0){20}}
\put(45,90){\line(-1,0){15}}
\put(65,150){\vector(0,-1){30}}
\put(65,120){\line(0,-1){30}}
\put(95,90){\vector(0,1){30}}
\put(95,120){\line(0,1){30}}
\put(170,65){\vector(-1,0){90}}
\put(80,65){\line(-1,0){50}}
\put(170,40){\vector(-1,0){90}}
\put(80,40){\line(-1,0){50}}
\end{picture}}
\put(250,360){
\begin{picture}(200,160)
\put(100,15){\makebox(0,0){(b2)}}
\put(170,90){\vector(-1,0){15}}
\put(155,90){\line(-1,0){20}}
\put(135,90){\vector(0,1){30}}
\put(135,120){\line(0,1){30}}
\put(105,150){\vector(0,-1){30}}
\put(105,120){\line(0,-1){30}}
\put(170,65){\vector(-1,0){50}}
\put(120,65){\line(-1,0){90}}
\put(170,40){\vector(-1,0){50}}
\put(120,40){\line(-1,0){90}}
\put(105,90){\vector(-1,0){50}}
\put(55,90){\line(-1,0){25}}
\multiput(70,65)(0,5){5}{\line(0,1){3}}
\end{picture}}
\put(250,180){
\begin{picture}(200,160)
\put(100,15){\makebox(0,0){(c2)}}
\put(170,90){\vector(-1,0){15}}
\put(155,90){\line(-1,0){20}}
\put(135,90){\vector(0,1){30}}
\put(135,120){\line(0,1){30}}
\put(105,150){\vector(0,-1){30}}
\put(105,120){\line(0,-1){30}}
\put(170,65){\vector(-1,0){50}}
\put(120,65){\line(-1,0){90}}
\put(170,40){\vector(-1,0){50}}
\put(120,40){\line(-1,0){90}}
\put(105,90){\line(-1,0){20}}
\multiput(55,90)(5,0){6}{\line(1,0){3}}

\put(55,90){\vector(-1,0){10}}
\put(45,90){\line(-1,0){15}}
\put(70,90){\oval(30,30)[b]}
\put(70,75){\vector(-1,0){0}}
\end{picture}}
\end{picture}}
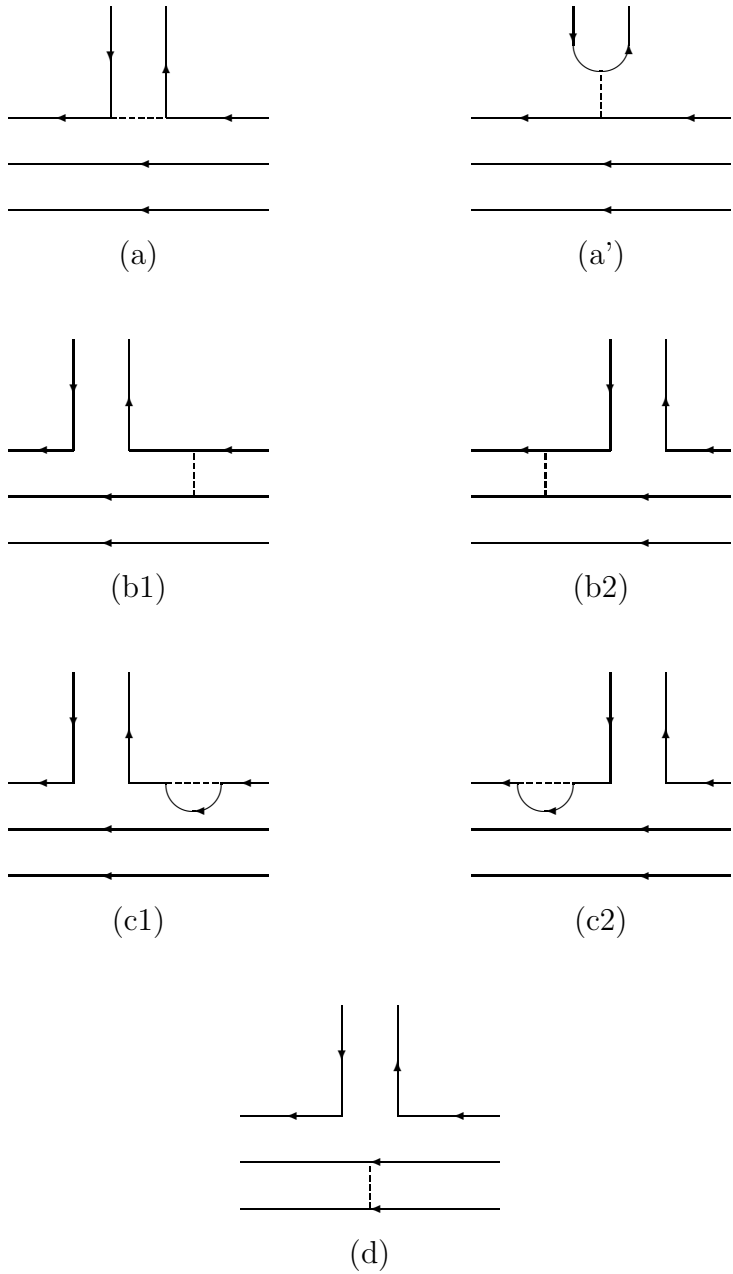
\put(235,10){\makebox(0,0)[b]{Fig.1. Quark diagrams for weak decays.}}
\end{picture}
\newpage
\begin{picture}(450,320)
\put(0,20){
\begin{picture}(200,130)
\put(30,40){\line(1,0){140}}
\put(30,65){\line(1,0){140}}
\put(30,90){\line(1,0){140}}
\multiput(100,65)(0,5){5}{\line(0,1){3}}
\put(30,72){\makebox(0,0){s}}
\put(30,97){\makebox(0,0){u}}
\put(170,72){\makebox(0,0){c}}
\put(170,97){\makebox(0,0){d}}
\end{picture}}
\put(80,150){\begin{picture}(90,40)
\put(20,20){\vector(0,1){20}}
\put(20,20){\vector(0,-1){20}}
\put(50,20){\makebox(0,0){SU(4)}}
\end{picture}}
\put(0,190){
\begin{picture}(200,130)
\put(30,40){\line(1,0){140}}
\put(30,65){\line(1,0){140}}
\put(30,90){\line(1,0){140}}
\multiput(100,65)(0,5){5}{\line(0,1){3}}
\put(30,72){\makebox(0,0){d}}
\put(30,97){\makebox(0,0){u}}
\put(170,72){\makebox(0,0){u}}
\put(170,97){\makebox(0,0){s}}
\end{picture}}
\put(250,190){
\begin{picture}(200,130)
\put(30,40){\line(1,0){140}}
\put(30,65){\line(1,0){140}}
\put(30,90){\line(1,0){55}}
\put(115,90){\line(1,0){55}}
\multiput(85,90)(5,0){6}{\line(1,0){3}}
\put(100,90){\oval(30,30)[b]}
\put(30,97){\makebox(0,0){d}}
\put(170,97){\makebox(0,0){s}}
\end{picture}}
\begin{picture}(450,20)
\put(100,5){\makebox(0,0){(1)}}
\put(350,5){\makebox(0,0){(2)}}
\end{picture}
\end{picture}
\\
\\
\\
\\
\baselineskip=10pt
Fig.2. Quark diagrams for the baryon-to-baryon matrix elements of the
parity conserving part of the weak Hamiltonian.
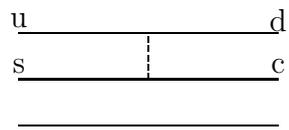
\end{document}